\documentclass[reprint,amsmath,amssymb,aps,pra,nofootinbib,superscriptaddress,showkeys]{revtex4-1}

\usepackage{graphicx}% Include figure files
\usepackage{dcolumn}% Align table columns on decimal point
\usepackage{bm}% bold math
\usepackage[table]{xcolor}
\usepackage{ulem}

\newcommand{\change}[1]{{\color{black}#1}}

\begin{document}

\title{Inner-shell excitation in the YbF molecule and its impact on laser cooling}

\author{Chi Zhang}
 \email{c.zhang@imperial.ac.uk}
\affiliation{Centre for Cold Matter, Blackett Laboratory, Imperial College London, Prince Consort Road, London SW7 2AZ UK}
\author{Chaoqun Zhang}
\affiliation{Department of Chemistry, The Johns Hopkins University, Baltimore, Maryland 21218, United States}
\author{Lan Cheng}
\email{lcheng24@jhu.edu}
\affiliation{Department of Chemistry, The Johns Hopkins University, Baltimore, Maryland 21218, United States}
\author{Timothy C. Steimle}
\affiliation{Department of Chemistry and Biochemistry, Arizona State University, Tempe, AZ 85287-1604, USA}
\author{Michael R. Tarbutt}
\email{m.tarbutt@imperial.ac.uk}
\affiliation{Centre for Cold Matter, Blackett Laboratory, Imperial College London, Prince Consort Road, London SW7 2AZ UK}

\begin{abstract}
The YbF molecule is a sensitive system for measuring the electron's electric dipole moment. The precision of this measurement can be improved by direct laser cooling of the molecules to ultracold temperature. However, low-lying electronic states arising from excitation of a 4f electron may hinder laser cooling. One set of these ``4f hole'' states lies below the $A^2\Pi_{1/2}$ excited state used for laser cooling, and radiative decay to these intermediate levels, even with branching ratios as small as $10^{-5}$, can be a hindrance. Other 4f hole states lie very close to the  $A^2\Pi_{1/2}$ state, and a perturbation results in states of mixed character that are involved in the laser cooling cycle. This perturbation may enhance the loss of molecules to states outside of the laser cooling cycle. We model the perturbation of the $A^2\Pi_{1/2}$ state to determine the strength of the coupling between the states, the de-perturbed potential energy curves, and the radiative branching ratios to various vibrational levels of the ground state, $X ^{2}\Sigma^+$. We use electronic structure calculations to characterise the 4f hole states and the strengths of transitions between these states and the $A^2\Pi_{1/2}$ and $X ^{2}\Sigma^+$ states. We identify a leak out of the cooling cycle with a branching ratio of roughly $5 \times 10^{-4}$, dominated by the contribution of the ground state configuration in a 4f hole state.
Finally, we assess the impact of these results for laser cooling of YbF and molecules with similar structure.
\end{abstract}


\maketitle

\section{Introduction}

Molecules containing heavy atoms can be sensitive probes of physics beyond the Standard Model \cite{Safronova2018}. For example, they can be used to measure the electron's electric dipole moment and to probe hadronic CP violating phenomena. Some of the molecules that have high sensitivity to new physics are also amenable to direct laser cooling and magneto-optical trapping. Such cooling and trapping may substantially improve the sensitivity of experiments by increasing coherence times, increasing  the number of useful molecules, and reducing systematic errors related to motion of the molecules. Good examples are YbF and YbOH; laser cooling of these species has been demonstrated~\cite{Lim2018,Alauze2021,Augenbraun2020} and experiments to use those cooled molecules to test fundamental physics are currently being constructed~\cite{Fitch2020b}. Cryogenic sources can produce beams of these molecules with speeds typically in the range 100--200~m/s~\cite{Hutzler2012}. They can then be decelerated to rest using radiation pressure and captured in a magneto-optical trap~\cite{Barry2014,Truppe2017b, Anderegg2017,Collopy2018}, provided leaks out of the laser cooling scheme can be reduced below about $10^{-5}$ so that at least $10^4$--$10^5$ photons can be scattered without excessive loss. For heavy molecules, the laser cooling scheme can be hindered by low-lying electronic states arising from excitation of inner-shell electrons. These states may lie close to the electronically excited state used for laser cooling, producing strong perturbations that can complicate the cooling scheme. Alternatively, they may lie below the state used for laser cooling, potentially introducing leaks out of the cooling cycle.  In this paper, we study these complications for the YbF molecule, which is an important molecule for measuring the electron's electric dipole moment~\cite{Hudson2011}.  Although we focus on YbF, the issues addressed here are likely to be common to many laser-coolable molecules containing heavy atoms.

\begin{figure}[tb]
\includegraphics[width=\columnwidth]{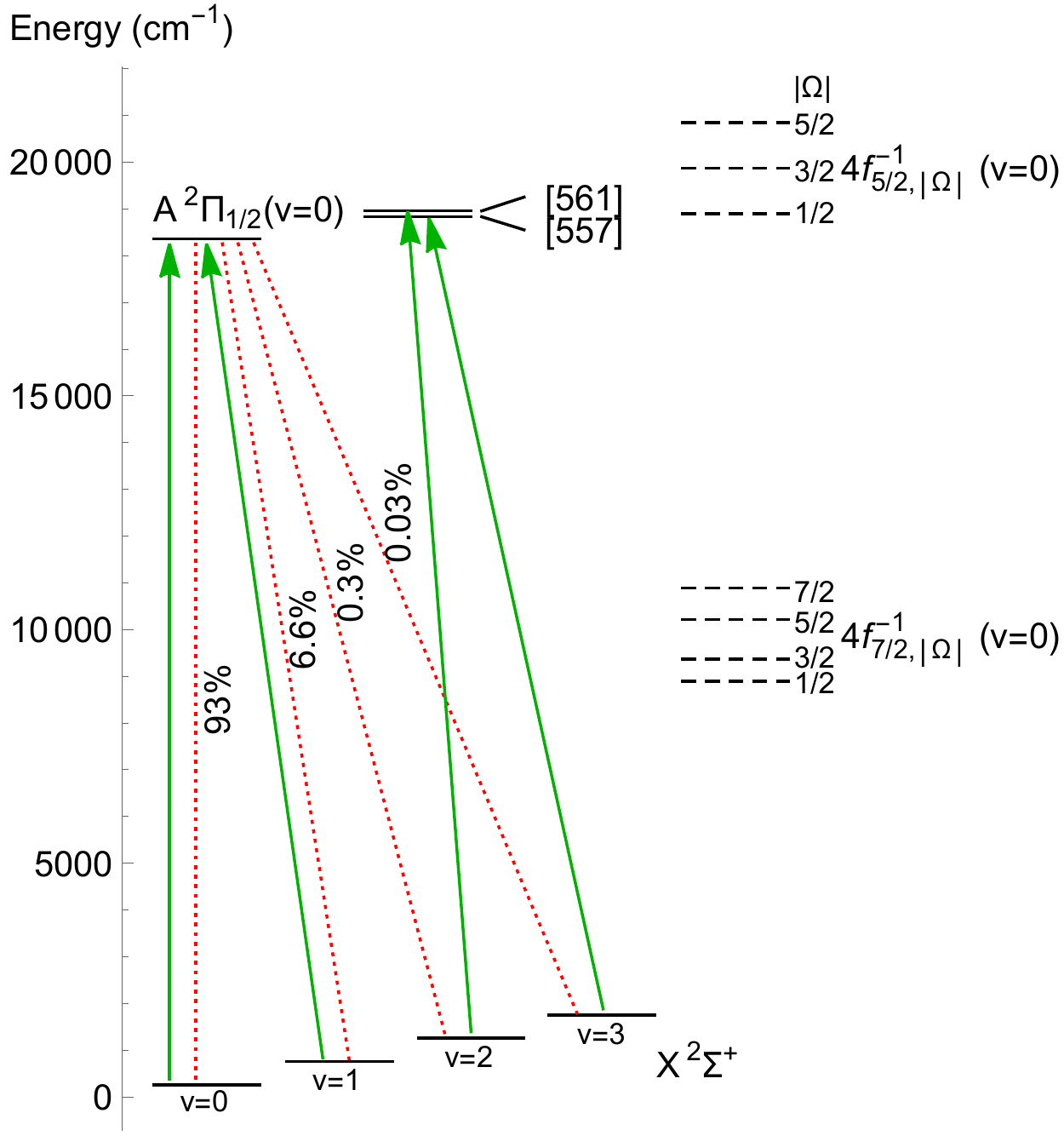}
\caption{Relevant energy levels of YbF and the scheme used for laser cooling. Solid (green) arrows show transitions driven by lasers. Dotted (red) lines show the vibrational branching ratios for decay of the $A{}^2\Pi_{1/2}(v=0)$ state. Dashed horizontal lines indicate approximate energies of states arising from excitation of a 4f electron. The states labelled [557] and [561] are mixtures of the $A{}^2\Pi_{1/2}(v=1)$ and $4f^{-1}_{5/2,1/2}(v=0)$ states. \label{fig:cooling_scheme}}
\end{figure}

Figure \ref{fig:cooling_scheme} illustrates the laser cooling scheme used for YbF~\cite{Smallman2014,Lim2018,Alauze2021}. The main laser cooling transition is the $A ^{2}\Pi_{1/2} \leftrightarrow X ^{2}\Sigma^{+}$ transition. The figure shows the vibrational branching ratios from $A ^{2}\Pi_{1/2} (v=0)$ and the laser transitions that are driven in order to address the $v=0,1,2,3$ vibrational levels of the ground state. In discussing the states shown in the figure, it is helpful to consider their long range character. The states $X ^{2}\Sigma^{+}$ and $A ^{2}\Pi$ correlate respectively to the $4f^{14}6s$ and $4f^{14}5d$ states of Yb$^+$. The dashed lines in the figure show levels arising from inner-shell excitation, correlating at long range to the $4f^{13}6s^2$ configuration of Yb$^{+}$. There are two groups of states due to a very large spin-orbit interaction, one correlating to $J=7/2$ and the other to $J=5/2$, with the former lying lower in energy. This spin-orbit splitting is about 10,000~cm$^{-1}$ in Yb$^{+}$. The electric field of the F$^-$ ion splits the $J=7/2$ state into four components with $|\Omega|=1/2,3/2,5/2,7/2$, and the $J=5/2$ state into three components with $|\Omega|=1/2,3/2,5/2$. Because they arise from a hole in the 4f shell, we label these states using the notation $4f^{-1}_{J,|\Omega|}$. There are currently no spectroscopic studies of these states, so their energies are uncertain and the positions indicated in Fig.~\ref{fig:cooling_scheme} are only approximate. As we will see later, the state $4f^{-1}_{5/2,1/2} (v=0)$ lies close to $A ^{2}\Pi_{1/2} (v=1)$ and the interaction between these states leads to eigenstates of mixed character which are labelled [557] and [561]\footnote{This notation specifies the energy above the ground state in THz.} in Fig.~\ref{fig:cooling_scheme}. The laser cooling scheme uses these mixed states for repumping population that reaches the $v=2$ and $v=3$ levels of $X ^{2}\Sigma^{+}$.

In this paper, we aim to address several questions relating to the $4f^{-1}_{J,|\Omega|}$ states, especially those that have an impact on laser cooling of YbF and similar molecules. The questions are as follows. What are the energies of these states? What is the strength of the coupling between the $A ^{2}\Pi_{1/2}$ and $4f^{-1}_{5/2,1/2}$ electronic states, what are the set of eigenstates produced by this coupling, and what are the branching ratios from each of these states to the various vibrational states of $X ^{2}\Sigma^{+}$? Can population in the excited states used for laser cooling decay to the $4f^{-1}_{7/2}$ states, and if so how large is this leak out of the cooling cycle? What is the lifetime of the $4f^{-1}_{7/2}$ states and which vibrational states of $X ^{2}\Sigma^{+}$ do they predominantly decay to? 
In Sect. \ref{depert} we use experimental data for vibrational levels to extract 
the diabatic potential energy curves of the $A ^{2}\Pi_{1/2}$ and $4f^{-1}_{5/2,1/2}$ states
and the electronic coupling between them. The branching ratios from these vibrational states to the vibrational levels of the ground state are calculated in Sect.~\ref{VibBran} using the derived potential energy curves and couplings, and compared with experimental values to verify the accuracy of the deperturbation procedure. 
{\it{Ab initio}} electronic and vibrational structure calculations are then presented in Sect. \ref{abini}
to provide an estimate of the level positions for the $4f^{-1}_{J,|\Omega|}$ states
and of the transition intensities for the $A^{2}\Pi_{1/2}\rightarrow 4f^{-1}_{7/2,1/2}$ transitions
and the $4f^{-1}_{7/2,1/2}\rightarrow X^2\Sigma^{+}$ transitions. The implications for laser cooling of YbF are discussed in Sect.~\ref{summary}.

\section{De-perturbation of $A ^{2}\Pi_{1/2}$ and $4f^{-1}_{5/2,1/2}$} \label{depert}

\begin{figure}[tb]
\includegraphics[width=\columnwidth]{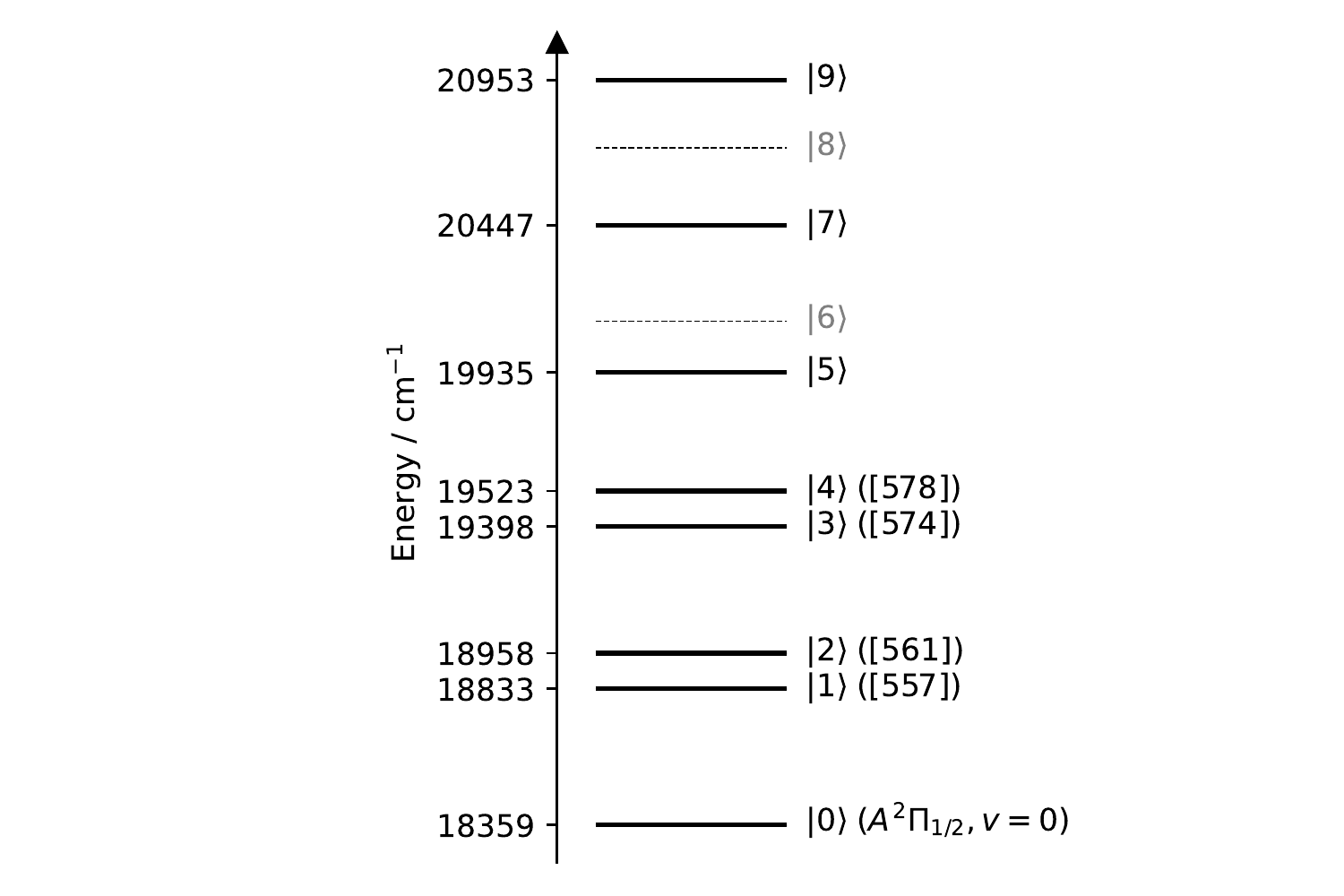}
\caption{\label{fig:measured_levels} Energies of excited states of YbF. Thick lines are observed states given by Refs.~\cite{Barrow1975,Lee1977, Dunfield1995}. All states are labeled by integers $|n\rangle$ in order of ascending energy. \change{Note that $n$ is just a label, not a vibrational quantum number.} The labels inside the parentheses are the standard notations of those states. For reference, in our convention (which differs from Refs.~ \cite{Barrow1975,Lee1977, Dunfield1995}), the energy of $X^{2}\Sigma^{+}(v=0)$ is 252.77~cm$^{-1}$. }
\end{figure}

Spectroscopic studies~\cite{Barrow1975,Lee1977, Dunfield1995} have established the energies of a set of vibrational levels between 18000 and 21000~cm$^{-1}$ above the ground state. These are shown in Fig.~\ref{fig:measured_levels}. The separations of the vibrational levels indicate that the first and second vibrationally excited states of $A^2\Pi_{1/2}$ are strongly perturbed. Theoretical~\cite{Dolg1992, Liu1998, Pototschnig21} and spectroscopic~\cite{Dunfield1995,Sauer1999, Lim2017} studies suggest that this is due to the interaction of the $A^2\Pi_{1/2}$ and $4f^{-1}_{5/2,1/2}$ states, which are close in energy and interact through a component of the electron-electron repulsion. \change{This perturbation only mixes states of the same $\Omega$, so we do not need to consider mixing with the $4f^{-1}_{5/2,3/2}$ and $4f^{-1}_{5/2,5/2}$ states~\cite{Lefebvre-Brion}}. In the following analysis we use the shorthand notation $|X\rangle$ for the $X^{2}\Sigma^{+}$ state, $|A\rangle$ for the $A^2\Pi_{1/2}$ state and $|4f\rangle$ for the $4f^{-1}_{5/2,1/2}$ state, and $H'$ for the perturbation. Vibrational levels are labelled $v_s$ with $s \in \{X, A,4f\}$. 

Our aim is to find the de-perturbed potential energy curves for the $|A\rangle$ and $|4f\rangle$ states. The energy of state $|s,v_s\rangle$ is described by $E_{s}(v)=T_{e,s}+\omega_{e,s}(v+1/2)-\omega_{e,s}x_{e,s}(v+1/2)^2$, where $T_e$ is the term energy, $\omega_e$ is the harmonic vibrational frequency, and $x_e$ accounts for the deviation from a harmonic oscillator potential. Their values for $|X\rangle$ are $T_{e,X}=0$, $\omega_{e,X}=506.67$~cm$^{-1}$ and $\omega_{e,X}x_{e,X}=2.25$~cm$^{-1}$~\cite{Dunfield1995}, which puts the energy of the lowest vibrational level of $|X\rangle$ at $E_X(0)=252.77$~cm$^{-1}$. In our model, $T_{e,A}$, $T_{e,4f}$, $\omega_{e,A}$, $\omega_{e,4f}$, $x_{e,A}$ and $x_{e,4f}$ are all free parameters. In addition, we introduce \change{a parameter $\Delta R$ which is the difference in the equilibrium internuclear separation for the two potentials}, and the parameter $\Omega = \langle A|H'|4f\rangle$ which is the coupling strength between the electronic states. This gives us 8 free parameters. We assume a separation of the electronic and vibrational degrees of freedom so that we can write $\langle A,v_A|H'|4f,v'_{4f}\rangle = \langle A|H'|4f\rangle \langle v_A|v'_{4f}\rangle = \Omega F_{vv'}$. Here,  $F_{v,v'}=\langle v_A|v'_{4f}\rangle$ is the overlap integral between vibrational wavefunctions.  Because we expect $x_e$ to be small, we use harmonic oscillator wavefunctions to approximate the low-lying vibrational wavefunctions so that $F_{vv'}$ can be calculated analytically. The effective Hamiltonian of the system in the $|s,v_s\rangle$ basis is
\begin{equation}
\label{bigH}
H=\sum_{s,v}E_{s}(v)|s,v_s\rangle\langle s,v_s| + \sum_{s\ne s',v,v'}\Omega F_{vv'}|s,v_s\rangle\langle s',v'_{s'}|.
\end{equation}
To the extent that this is an accurate representation of the problem, all 8 parameters can be determined by fitting the eigenvalues of $H$ to the measured energy levels shown in Fig.~\ref{fig:measured_levels}.

We make a first estimate of the parameters as follows.  We note that the $v_A=0,3,4,5$ states of $|A\rangle$ are not strongly perturbed, and we use these states to set the initial values $T_{e,A}\rightarrow 18103.4~\mathrm{cm}^{-1}$, $\omega_{e,A}\rightarrow 532.4~\mathrm{cm}^{-1}$ and $x_{e,A}\omega_{e,A} \rightarrow 2.7$~cm$^{-1}$. These parameters give us our initial estimates of the energies of the unperturbed $v_A=1,2$ levels. We find that the perturbation shifts them by $-62.8~\mathrm{cm}^{-1}$ and $-20.2~\mathrm{cm}^{-1}$, respectively. Since $v_A=0$ is unperturbed, while $v_A=1,2$ are strongly perturbed, we conclude that the energy of $v_{4f}=0$ must be close to $v_A=1$. Next, for our initial estimates, we suppose that the main contribution to the shifts of each of the $|A\rangle$ states comes from interaction with the closest vibrational state of $|4f\rangle$. At this level of approximation, the perturbation shifts pairs of levels, one from $|A\rangle$ and one from $|4f\rangle$, by equal and opposite amounts. Applying this two-state model separately to the $\{|A,1_A\rangle,|4f,0_{4f}\rangle\}$ subsystem and the $\{|A,2_A\rangle,|4f,1_{4f}\rangle\}$ subsystem we obtain $\omega_{e,4f} \approx  606~\mathrm{cm}^{-1}$,  $T_{e,4f} \approx 18598~\mathrm{cm}^{-1}$, $\Omega \langle 1_A|0_{4f}\rangle \approx 62~\mathrm{cm}^{-1}$ and $\Omega \langle 2_A|1_{4f}\rangle \approx 44~\mathrm{cm}^{-1}$. The ratio of the last two parameters gives $\langle 1_A|0_{4f}\rangle / \langle 2_A|1_{4f}\rangle = 1.41$. For harmonic oscillator wavefunctions, this ratio of overlap integrals depends only on $\omega_{e,A}$, $\omega_{e,4f}$, and $\Delta R$. The first two are already estimated, so the calculation gives $\Delta R$. We find $\Delta R=1.99a_A$ where $a_A=\sqrt{\frac{\hbar}{2m\omega_{e,A}}}$ is the harmonic oscillator length of the $|A\rangle$ state. Knowing this, we obtain the electronic part of the coupling coefficient to be $\Omega \approx 99~\mathrm{cm}^{-1}$.

\begin{figure}[tb]
\includegraphics[width=\columnwidth]{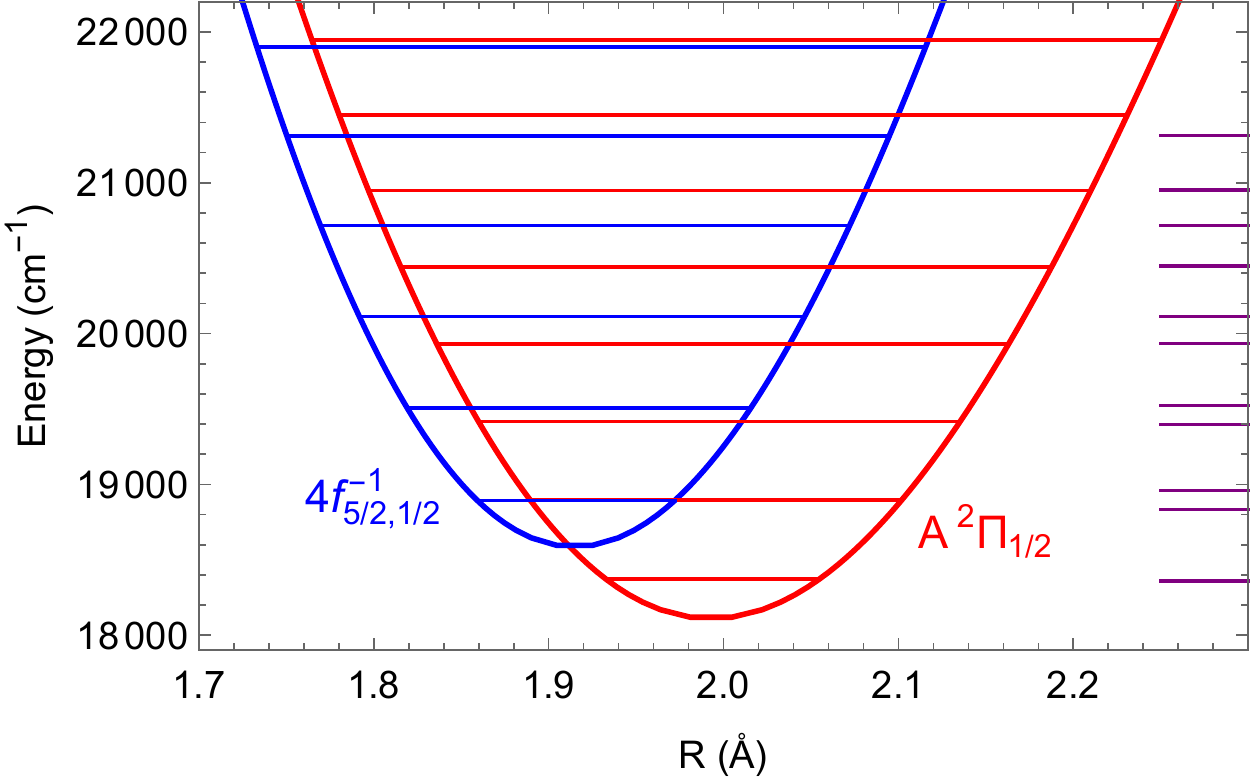}
\caption{Low-lying parts of the potential energy curves for the $A ^{2}\Pi_{1/2}$ and $4f^{-1}_{5/2,1/2}$ states resulting from the de-perturbation procedure. The vibrational states within each potential are also shown. The levels on the right hand side show the set of levels produced by the coupling between the two states. By construction, the levels match the 8 that have been observed spectroscopically (shown in Fig. \ref{fig:measured_levels}). \label{fig:deperturbed_potentials}}
\end{figure}

With the help of these initial estimates, we are able to fit the eigenvalues of Hamiltonian (\ref{bigH}) to the set of measured levels (see Fig.~\ref{fig:measured_levels}) to obtain a set of refined parameters. The best fit parameters are listed in Table~\ref{parameters}. The root mean square deviation between all calculated eigenenergies and the experimental data is  $0.26~\mathrm{cm}^{-1}$, more than three orders of magnitude smaller than the typical vibrational level spacing \change{and approximately equal to the uncertainties in the experimental data~\cite{Barrow1975,Lee1977, Dunfield1995}}. Figure \ref{fig:deperturbed_potentials} shows the de-perturbed potentials obtained from the procedure described above, together with the vibrational states within each potential. The eigenvalues of Eq.~(\ref{bigH}) are shown on the right hand side of the figure.  Table~\ref{eigenstates} gives the corresponding eigenstates of Hamiltonian (\ref{bigH}). Here, we have used $|n\rangle$ to label the states in order of ascending energy. $|1\rangle$, $|2\rangle$, $|3\rangle$ and $|4\rangle$ correspond to the perturbed states [557], [561], [574] and [578], respectively.

\begin {table*}
\caption {Best fit parameters obtained from the de-perturbation procedure}
\label{parameters} 
\begin{center}
\begin{tabular}{ c|c|c|c|c|c|c|c|c } 
\hline
parameter & $T_{e,A}$ & $\omega_{e,A}$ & $\omega_{e,A}x_{e,A}$ & $T_{e,4f}$ & $\omega_{e,4f}$ & $\omega_{e,4f}x_{e,4f}$ & $\Omega$ &$\Delta R$ \\
\hline
value & $18109.4~\mathrm{cm}^{-1}$ & $528.7~\mathrm{cm}^{-1}$ & $2.27~\mathrm{cm}^{-1}$ & $18585.7~\mathrm{cm}^{-1}$ & $618.9~\mathrm{cm}^{-1}$ & $2.91~\mathrm{cm}^{-1}$ & $108.3~\mathrm{cm}^{-1}$ & $7.79~\mathrm{pm}$ \\ 
\hline
\end{tabular}
\end{center}
\end {table*}

\section{Branching ratios to the vibrational levels of $X{}^{2}\Sigma^{+}$} \label{VibBran}

\begin {table*}
\caption {Eigenstates of the full Hamiltonian in the $\{|A,v_A\rangle, |4f,v_{4f}\rangle\}$ basis. The shading highlights coefficients with significant amplitudes.}
\label{eigenstates} 
\begin{center}
\begin{tabular}{c|c|c|c|c|c|c||c|c|c|c|c} 
\hline
 & $|A,0_A\rangle$ & $|A,1_A\rangle$ & $|A,2_A\rangle$ & $|A,3_A\rangle$ & $|A,4_A\rangle$ & $|A,5_A\rangle$ & $|4f,0_{4f}\rangle$ & $|4f,1_{4f}\rangle$ & $|4f,2_{4f}\rangle$ & $|4f,3_{4f}\rangle$ & $|4f,4_{4f}\rangle$ \\
\hline
$|0\rangle$ & \cellcolor{blue!30}-0.989 & 0.011 & -0.003 & 0.0 & 0.0 & 0.0 & \cellcolor{blue!10}0.130 & 0.058 & 0.024 & 0.008 & 0.002 \\ 
\hline
$|1\rangle$ & -0.095 & \cellcolor{blue!20}-0.723 & 0.052 & -0.016 & 0.006 & -0.002 & \cellcolor{blue!20}-0.679 & 0.023 & 0.035 & 0.021 & 0.008 \\ 
\hline
$|2\rangle$ & 0.087 & \cellcolor{blue!20}-0.687 & -0.071 & 0.022 & -0.008 & 0.003 & \cellcolor{blue!20}0.716 & -0.009 & 0.030 & 0.023 & 0.011 \\ 
\hline
$|3\rangle$ & 0.032 & -0.006 & \cellcolor{blue!30}0.916 & -0.031 & 0.011 & -0.004 & 0.087 & \cellcolor{blue!20}0.384 & 0.029 & -0.025 & -0.027 \\ 
\hline
$|4\rangle$ & 0.052 & 0.012 & \cellcolor{blue!20}-0.385 & \cellcolor{blue!10}-0.112 & 0.041 & -0.018 & -0.016 & \cellcolor{blue!30}0.912 & -0.030 & 0.006 & 0.014 \\ 
\hline
$|5\rangle$ & -0.007 & -0.011 & 0.015 & \cellcolor{blue!30}-0.986 & -0.0 & 0.0 & 0.026 & \cellcolor{blue!10}-0.117 & \cellcolor{blue!10}-0.103 & -0.050 & 0.008 \\ 
\hline
$|6\rangle$ & -0.024 & -0.046 & 0.038 & \cellcolor{blue!10}0.105 & \cellcolor{blue!10}0.126 & -0.051 & 0.002 & -0.008 & \cellcolor{blue!30}-0.983 & 0.012 & 0.004 \\ 
\hline
$|7\rangle$ & -0.002 & -0.004 & -0.001 & 0.007 & \cellcolor{blue!30}-0.990 & -0.012 & -0.009 & 0.041 & \cellcolor{blue!10}-0.126 & 0.014 & -0.047 \\ 
\hline
$|8\rangle$ & -0.008 & -0.031 & -0.027 & 0.051 & -0.013 & \cellcolor{blue!10}0.120 & -0.001 & 0.005 & -0.014 & \cellcolor{blue!30}-0.990 & -0.002 \\ 
\hline
$|9\rangle$ & 0.0 & 0.001 & 0.0 & -0.003 & -0.0 & \cellcolor{blue!30}0.990 & -0.004 & 0.017 & -0.050 & \cellcolor{blue!10}0.120 & -0.056 \\ 
\hline
$|10\rangle$ & 0.002 & 0.014 & 0.031 & 0.009 & -0.047 & 0.055 & -0.0 & 0.001 & -0.004 & 0.005 & \cellcolor{blue!30}0.997 \\ 
\hline
\end{tabular}
\end{center}
\end {table*}

Next, we calculate the vibrational branching ratios from each of the states  $|n\rangle$ to each of the ground states $|X,v_X\rangle$. The transition dipole moments $\langle n|d|X,v_X\rangle$  can be expressed in terms of $\langle A|d|X\rangle$, $\langle 4f|d|X\rangle$ and the Franck-Condon factors of the $|A\rangle\leftrightarrow|X\rangle$ and $|4f\rangle\leftrightarrow|X\rangle$ transitions. Since $|\langle 4f|d|X\rangle|\ll |\langle A|d|X\rangle|$ \change{(see later, especially Table \ref{f13tms})}, the contribution of $|4f\rangle$ to the transition dipole moments can be neglected and all decay rates $|n\rangle\rightarrow|X,v_X\rangle$ are proportional to $|\langle A|d|X\rangle|^2$. As a consequence, the radiative branching ratios can be determined exclusively from the Franck-Condon factors of the $|A\rangle\leftrightarrow|X\rangle$ transitions, which can be calculated using the harmonic oscillator wavefunctions with a separation of $2.1~\mathrm{pm}$. The calculated branching ratios are listed in Table.~\ref{branching} and compared with experimental data from ref. \cite{Smallman2014} and \cite{Zhuang2011}.

\begin {table*}
\caption {Radiative branching ratios to $|X{}^{2}\Sigma^{+},v_X\rangle$. Where there are two entries, the first gives our calculated value and the second gives the experimental values from Refs. \cite{Smallman2014}  and \cite{Zhuang2011}. The values for states $|6\rangle$, $|8\rangle$ and $|10\rangle$ are not given because these are mainly of $|4f\rangle$ character and their branching ratios may be significantly influenced by even a small (non-zero) value of $\langle 4f | d | X \rangle$.}
\label{branching} 
\begin{center}
\begin{tabular}{c|c c c c c c c c} 
\hline
 & $v_X=0$ & $v_X=1$ & $v_X=2$ & $v_X=3$ & $v_X=4$ & $v_X=5$ & $v_X=6$ & $v_X=7$ \\
\hline
$|0\rangle \approx |A^{2}\Pi_{1/2},v_A=0\rangle$ & $93.6$ & $6.0$ & $0.4$ & $0.02$ & $0.001$ & $<10^{-5}$ & $<10^{-5}$ & $<10^{-5}$ \\ 

 &  $93.3 \pm 0.3$ & $6.6 \pm 0.3$ & $0.3 \pm 0.05$ & $<0.01$ \\ 
\hline
$|1\rangle \equiv$ [557] & 15.1 & 70.2 & 12.8 & 1.6 & 0.2 & 0.02 & 0.003 & $<10^{-5}$ \\ 

 &  $13.2 \pm 0.5$ & $70.7 \pm 0.6$ & $13.9 \pm 0.2$ & $1.9 \pm 0.2$ & $<0.2$ \\ 
\hline
$|2\rangle \equiv$ [561] & 1.9 & 91.6 & 6.4 & 0.1 & 0.005 & 0.009 & 0.002 & $<10^{-5}$ \\ 

 &  $2.8 \pm 0.3$ & $89 \pm 1$ & $7.8 \pm 0.2$ & $< 0.2$ \\ 
\hline
$|3\rangle \equiv$ [574] & 0.5 & 12.0 & 69.4 & 15.7 & 2.1 & 0.2 & 0.02 & 0.002 \\ 

 &  $<1$ & $8.1 \pm 0.6$ & $75 \pm 2$ & $15 \pm 2$ & $1.9\pm 0.8$ \\ 
\hline
$|4\rangle \equiv$ [578]  & 1.2 & 13.0 & 81.4 & 3.5 & 0.1 & 0.6 & 0.15 & 0.02 \\ 

 &  $<1$ & $13.3 \pm 0.6$ & $82.5 \pm 0.7$ & $4.1 \pm 0.6$ & $<1$ \\ 
\hline
$|5\rangle$  & 0.01 & 0.4 & 17.4 & 62.2 & 17.2 & 2.4 & 0.2 & 0.01 \\ 
\hline
$|7\rangle$  & $<10^{-5}$ & 0.003 & 0.6 & 23.5 & 53.6 & 18.8 & 3.2 & 0.3 \\ 
\hline
$|9\rangle$  & $<10^{-5}$ & $<10^{-5}$ & $<10^{-5}$ & 1.1 & 28.8 & 44.6 & 21.1 & 4.4 \\ 
\hline
\end{tabular}
\end{center}
\end {table*}

In most cases, our calculated branching ratios are consistent with the experimental measurements. This good agreement supports our descriptions of the perturbed states. In particular, we note that our model correctly predicts most of the branching ratios from [557], [561], [574] and [578]; these values were not understood prior to this work. It is worth noting that the interference between different vibrational states can alter the branching ratios significantly. For instance, as can be seen from Table~\ref{eigenstates}, [557] and [561] states have very similar $|1_A\rangle$ probabilities (52\% for [557], 47\% for [561]), and less than 1\% probability for any other $|v_A\ne 1\rangle$ state. The state $|1_A\rangle$ decays mainly to $|1_X\rangle$, and yet the branching ratio to $|1_X\rangle$ from [561] is about 20\% higher than that from [557]. This may be understood as constructive or destructive interferences between different decay processes, depending on the relative phase of different vibrational components, or equivalently, that the main vibrational wavefunction $|1_A\rangle$ is dressed differently by other vibrational components depending on their phases.

The small differences between our calculated branching ratios and the measured values might be due to deviations from harmonic oscillator wavefunctions (though we note that a calculation using Rydberg–Klein–Rees [RKR] potentials makes very little difference), or could be the influence of other perturbing states that are not included in our model.

\section{Electronic and vibrational structure calculations} \label{abini}

The {\it{ab initio}} calculations carried out here aim to estimate the energy levels of the $4f^{-1}_{J,|\Omega|}$ states and the intensities of the $A ^{2}\Pi_{1/2}\leftrightarrow 4f^{-1}_{J,|\Omega|}$
and $4f^{-1}_{J,|\Omega|} \leftrightarrow X^2\Sigma^{+}$ transitions. 
\change{In general it is a formidable task to compute relative energies of molecular states to an accuracy
within 100 cm$^{-1}$ \cite{Tajti04,Karton06}.
In the present calculations, in addition to intrinsic errors of approximate electron-correlation methods,}
a challenge in accurate calculations of the term energies for the $4f^{-1}_{J,|\Omega|}$ states 
lies in the sensitivity of computed results to basis-set effects. 
Since a $4f^{-1}_{J,|\Omega|}$ state differs from the $X^2\Sigma^{+}$ state 
by an excitation from a Yb $4f$ orbital to a Yb $6s$ orbital, i.e., 
an excitation involving two orbitals with very different angular momenta, 
the computed energies exhibit large differential basis-set effects. 
We address this challenge using the synergy between experimental analysis and computations; 
we obtained the relative energies between the $4f^{-1}_{J,|\Omega|}$ states from electronic-structure calculations 
and then used the experimental term energy of the $4f^{-1}_{5/2,1/2}$ state obtained 
in Sec.~\ref{depert} to derive the term energies for the other $4f^{-1}_{J,|\Omega|}$ states.

To account for relativistic and electron-correlation effects in YbF, 
we have performed relativistic exact two-component (X2C) \cite{Dyall97,Ilias07,Liu09} coupled-cluster singles and doubles (CCSD) \cite{Purvis82,Liu18b} calculations for the $X^2\Sigma^{+}$ state and equation-of-motion CCSD (EOM-CCSD) \cite{Stanton93a,Asthana19} calculations for the excited states. 
Scalar-relativistic effects and spin-orbit coupling have been included in the orbitals non-perturbatively 
using an X2C Hamiltonian with atomic mean-field spin-orbit integrals (X2CAMF) \cite{Liu18}. The Breit term has been included in the AMF scheme. We have used the CFOUR program package \cite{Matthews20CFOUR,cfour,Liu21a} for all calculations
presented here. 

Since we would like to calculate transition dipole moments for the $A ^{2}\Pi_{1/2}\leftrightarrow 4f^{-1}_{J,|\Omega|}$
%and $X^2\Sigma_{1/2}\leftrightarrow 4f^{-1}_{J,|\Omega|}$ 
transitions, the $A ^{2}\Pi_{1/2}$ and $4f^{-1}_{J,|\Omega|}$ states need to be treated on the same footing
in X2CAMF-EOM-CCSD calculations. We used an electronic ground state $X^2\Sigma^{+},\Omega=1/2$ of YbF
%, with an approximate Yb atomic configuration 4f146s1, 
as the reference state in EOM-CCSD calculations 
and obtained the targeted electronically excited states by applying excitation operators on this ground-state wave function. As shown in Table \ref{excite}, the $A ^{2}\Pi_{1/2}$ states 
%with an approximate Yb atomic configuration $4f^{14}6p^{1}_{1/2}$ 
are obtained by exciting an unpaired electron from %a $\sigma_{1/2}$ orbital of 
a Yb 6s orbital into 
%a $\pi_{1/2}$ orbital of 
a Yb $6p_{1/2}$ orbital. The $4f^{-1}_{J,|\Omega|}$ states are obtained by exciting a Yb $4f$ electron into the unoccupied Yb $6s$ orbital. In this way, the excitation operators to obtain the $A ^{2}\Pi_{1/2}$ and $4f^{-1}_{J,|\Omega|}$ states are all dominated by single excitations that can be accurately calculated using the EOM-CCSD method.  

\begin {table*}
\caption {Excitations from an electronic ground state $X^2\Sigma_{1/2},\Omega=1/2$ wave function
to obtain the targeted $A ^{2}\Pi_{1/2}$ and $4f^{-1}_{J,|\Omega|}$ states.}
\begin{center}
\begin{tabular}{c|c} 
\hline
 Nominal Excitations & Excited States  \\
\hline
Yb $6s_{1/2}~,~j_z=1/2$ $\to$ Yb $6p_{1/2}~,~j_z=\pm 1/2$ ~ ~& ~ ~
$A^2\Pi_{1/2}~,~\Omega=\pm 1/2$ \\
Yb $4f_{7/2}~,~j_z=\pm 1/2$ $\to$ Yb $6s_{1/2}~,~j_z=-1/2$ ~ ~& ~ ~
$4f^{-1}_{7/2,1/2}~,~ \Omega=\mp 1/2$ \\
Yb $4f_{7/2}~,~j_z=\pm 3/2$ $\to$ Yb $6s_{1/2}~,~j_z=-1/2$ ~ ~& ~ ~
$4f^{-1}_{7/2,3/2}~,~ \Omega=\mp 3/2$ \\
Yb $4f_{7/2}~,~j_z=\pm 5/2$ $\to$ Yb $6s_{1/2}~,~j_z=-1/2$ ~ ~& ~ ~
$4f^{-1}_{7/2,5/2}~,~ \Omega=\mp 5/2$ \\
Yb $4f_{7/2}~,~j_z=\pm 7/2$ $\to$ Yb $6s_{1/2}~,~j_z=-1/2$ ~ ~& ~ ~
$4f^{-1}_{7/2,7/2}~,~ \Omega=\mp 7/2$ \\
Yb $4f_{5/2}~,~j_z=\pm 1/2$ $\to$ Yb $6s_{1/2}~,~j_z=-1/2$ ~ ~& ~ ~
$4f^{-1}_{5/2,1/2}~,~ \Omega=\mp 1/2$ \\
Yb $4f_{5/2}~,~j_z=\pm 3/2$ $\to$ Yb $6s_{1/2}~,~j_z=-1/2$ ~ ~& ~ ~
$4f^{-1}_{5/2,3/2}~,~ \Omega=\mp 3/2$ \\
Yb $4f_{5/2}~,~j_z=\pm 5/2$ $\to$ Yb $6s_{1/2}~,~j_z=-1/2$ ~ ~& ~ ~
$4f^{-1}_{5/2,5/2}~,~ \Omega=\mp 5/2$ \\
\hline
\end{tabular}
\label{excite}
\end{center}
\end {table*}

We have used two basis sets for Yb in our calculations. One is the uncontracted version of the ANO-RCC (ANO-RCC-unc) basis set of triple-$\zeta$ (TZ) quality \cite{Roos08a}. The combination of the Yb ANO-RCC-unc set with the F uncontracted aug-cc-pVTZ basis set \cite{Dunning89,Kendall92a} will be denoted as “TZ” in the following discussions. 
%The other basis set for Yb is of quadruple-zeta quality, with 
We have also augmented the Yb ANO-RCC-unc set with g-, h-, and i-type functions from the correlation-consistent quadruple-$\zeta$ basis set \cite{Lu16} to obtain a Yb basis set of quadruple-$\zeta$ (QZ) quality. The combination between this Yb basis set and the uncontracted aug-cc-pVTZ for F will be denoted as “QZ”. 
All CC calculations have kept frozen 48 core electrons and virtual orbitals higher than 100 Hartree.

We have computed local potential energy curves for $X^2\Sigma^{+}$, $A ^{2}\Pi_{1/2}$,
and $4f^{-1}_{J,|\Omega|}$ states \change{in the range of 
1.84 {\AA} to 2.18 {\AA} with an interval of 0.02 {\AA}.}
We have fit \change{seven data points on the surface for each state around the corresponding equilibrium
bond length, i.e., the data points ranging from 1.96 {\AA} to 2.08 {\AA} for the $X^2\Sigma^{+}$ state, 1.94 {\AA} to 2.06 {\AA} for the $A ^{2}\Pi_{1/2}$ state, 
and 1.88 {\AA} to 2.00 {\AA} for the $4f^{-1}_{J,|\Omega|}$ states}, to fourth-order polynomial functions,
and extracted Morse potential functions from the equilibrium bond lengths, equilibrium electronic energies,
and force constants obtained in the fitting procedure. 
The spectroscopic constants were then obtained using these potentials
and vibrational second-order perturbation (VPT2) \cite{Mills72} theory. 
We have fit the computed electronic transition dipole moments to quadratic functions
with respect to bond lengths and have carried out
discrete variable representation (DVR)
\cite{Colbert92} calculations for vibrational wave functions using
X2CAMF-EOM-CCSD/TZ Morse potential functions.
We have then computed transition dipole moments
for vibronic wave functions using these transition dipole functions
and vibrational wave functions.

\subsection{Spectroscopic constants for the $4f^{-1}_{J,|\Omega|}$ states}

\begin {table*}
\caption {Spectroscopic constants for the $X^2\Sigma^{+}$ state obtained from X2CAMF-CCSD calculations and those for $4f^{-1}_{J,|\Omega|}$ states and the $A ^{2}\Pi_{1/2}$ state from X2CAMF-EOM-CCSD calculations. }
\begin{center}
\begin{tabular}{cc|ccccc} 
\hline
 & &  & &  & \multicolumn{2}{c}{$T_e$ (cm$^{-1}$)} \\
 & & ~ $R_e$ (\AA) ~ & ~ $\omega_e$ (cm$^{-1}$) ~
 & ~ $\omega_e x_e$ (cm$^{-1}$) ~ & computed & ~ combined$^\text{c}$ ~ \\
\hline
~~TZ~~ & $4f^{-1}_{5/2,5/2}$ & --$^\text{b}$ & --$^\text{b}$  & --$^\text{b}$ & 18018     & 20543            \\
       & $4f^{-1}_{5/2,3/2}$ & 1.936         & 603.4          & 3.8           & 17040     & 19565            \\
       & $4f^{-1}_{5/2,1/2}$ & 1.928         & 612.6          & 3.4           & 16060     & 18586$^\text{a}$ \\
       & $4f^{-1}_{7/2,7/2}$ & 1.936         & 600.3          & 2.7           & 8059      & 10584            \\
       & $4f^{-1}_{7/2,5/2}$ & 1.936         & 605.2          & 2.7           & 7383      & 9908            \\
       & $4f^{-1}_{7/2,3/2}$ & 1.932         & 610.3          & 2.7           & 6539      & 9064             \\
       & $4f^{-1}_{7/2,1/2}$ & 1.931         & 619.3          & 2.5           & 6054      & 8580             \\
       & $A ^{2}\Pi_{1/2}$   & 2.001         & 528.2          & 2.2           & 18332     & 18109$^\text{a}$ \\
       & $X^2\Sigma_{1/2}$   & 2.028         & 500.2          & 2.2           & 0         & 0                \\ % 500.2 & 2.2
\hline
~~QZ~~ & $4f^{-1}_{5/2,5/2}$ & --$^\text{b}$ & --$^\text{b}$  & --$^\text{b}$ & 18770 & 20557 \\ 
& $4f^{-1}_{5/2,3/2}$ & 1.936 & 602.8  & 3.6 &  17787 & 19574 \\
& $4f^{-1}_{5/2,1/2}$ & 1.928 & 612.3  & 3.1 & 16799 & 18586$^\text{a}$ \\
& $4f^{-1}_{7/2,7/2}$ & 1.936 & 600.6  & 2.7 & 8792 & 10579 \\
& $4f^{-1}_{7/2,5/2}$ & 1.936   & 605.4 & 2.7 & 8120 & 9907\\
& $4f^{-1}_{7/2,3/2}$ & 1.931 & 610.6  & 2.7 & 7274 & 9061\\ 
& $4f^{-1}_{7/2,1/2}$ & 1.929 & 618.3  & 2.7 & 6764 & 8551 \\
& $A ^{2}\Pi_{1/2}$   & --$^\text{b}$ & --$^\text{b}$  & --$^\text{b}$ & 18431 & 18109$^\text{a}$\\
& $X^2\Sigma_{1/2}$   & 2.028 & 501.8  & 2.2 & 0 & 0 \\
\hline
\end{tabular}
\label{f13ene}
\end{center}
a. fixed to the values derived in Sect.~\ref{depert} \\
b. small discontinuity in the computed curve because of perturbation by a nearby state.\\
c. derived by shifting the computed energies of all the $4f^{-1}$ states so that $T_e$ for the $4f^{-1}_{5/2,1/2}$ state matches the value determined from the de-perturbation analysis in Sect.~\ref{depert}.
\end {table*}

Table \ref{f13ene} gives the spectroscopic constants obtained from the calculations. The computed bond lengths of the $4f^{-1}_{J,|\Omega|}$ states amount to 1.93-1.94 {\AA}, substantially shorter than those of 2.00 {\AA}
%Å 
for the $A ^{2}\Pi_{1/2}$ state and 2.03 {\AA} for the $X^2\Sigma^{+}$ state. 
This might be attributed partly to the attraction between the Yb 4f core hole
and the partially negatively charged fluorine site and partly to having one more electron in a back-polarized Yb 6s orbital reducing
the electrostatic repulsion with the electrophilic fluorine center.
The computed harmonic vibrational frequencies of 600-620 cm$^{-1}$ for the $4f^{-1}_{J,|\Omega|}$ states are larger than the values of 528 cm$^{-1}$ for the $A ^{2}\Pi_{1/2}$ state and 502 cm$^{-1}$ for the $X^2\Sigma^{+}$ state. The harmonic frequencies within the $4f^{-1}_{J,|\Omega|}$ manifold tend to decrease slightly with increasing $|\Omega|$. For example, the harmonic frequencies of the $4f^{-1}_{7/2,1/2}$, $4f^{-1}_{7/2,3/2}$, $4f^{-1}_{7/2,5/2}$, and $4f^{-1}_{7/2,7/2}$ states take the values of 619 cm$^{-1}$, 610 cm$^{-1}$, 605 cm$^{-1}$, and 600 cm$^{-1}$, respectively. This is consistent with the observation that the bond lengths tend to increase slightly along this series.
The computed properties for the ground state and the $A ^{2}\Pi_{1/2}$ state compare favorably with the experimental data.

As shown in Table \ref{f13ene}, the differences between the TZ and QZ results for the computed term energies for the $4f^{-1}_{J,|\Omega|}$ states amount to more than 700 cm$^{-1}$. In contrast, the relative shifts of energies between two $4f^{-1}_{J,|\Omega|}$ states are relatively insensitive to basis-set effects. Using the EOM-CCSD/QZ relative energies among the $4f^{-1}_{J,|\Omega|}$ states together with the “experimental” value of 18586 cm$^{-1}$ for the term energy of the $4f^{-1}_{5/2,1/2}$ state derived in Sect. \ref{depert},
we estimate the term energies of all the other $4f^{-1}_{J,|\Omega|}$ states. {These are the values given in the last column of Table~\ref{f13ene}, labelled ``combined''.}
For example, the equilibrium term energies for the $4f^{-1}_{7/2,1/2}$ state and the $4f^{-1}_{7/2,3/2}$ state are estimated to be 8551~cm$^{-1}$ and 9061~cm$^{-1}$, respectively. 

The computed spectroscopic constants for the $4f^{-1}_{5/2,1/2}$ and $A ^{2}\Pi_{1/2}$ states agree {remarkably} well with those derived for the diabatic electronic states in Sect. \ref{depert}. For both states, the computed bond lengths and harmonic frequencies are all within 1\% of the values found from the deperturbation analysis. Note that the electronic-structure calculations solve the electronic eigenvalue
equation and produce wave functions and energies for adiabatic electronic states. 
In principle the calculations should capture the electronic couplings between the $4f^{-1}_{5/2,1/2}$ and $A ^{2}\Pi_{1/2}$ states discussed in the Sect. \ref{depert}. However, because of the differential basis-set effects, our X2CAMF-EOM-CCSD/TZ calculations underestimate the electronic energies of the $4f^{-1}_{5/2,1/2}$ states relative to the $A ^{2}\Pi_{1/2}$ states by around 2500 cm$^{-1}$. Consequently, these calculations do not reproduce the near-degeneracy between the $A ^{2}\Pi_{1/2}, v=1$ state and the $4f^{-1}_{5/2,1/2}, v=0$ state. Instead, the computed local potential energy curves for the $4f^{-1}_{5/2,1/2}$ and $A ^{2}\Pi_{1/2}$ states agree well with those derived for the diabatic electronic states. The X2CAMF-EOM-CCSD/QZ results exhibit an underestimation of electronic energies of the $4f^{-1}_{5/2,1/2}$ state relative to $A ^{2}\Pi_{1/2}$ by around 1800 cm$^{-1}$. For this reason, instead of having a crossing around 1.91 {\AA} as shown in Fig. \ref{fig:deperturbed_potentials}, the crossing in the calculation lies around 2.05 {\AA}. The corresponding $A ^{2}\Pi_{1/2}$ curve thus shows small discontinuities and could not be fit to obtain spectroscopic constants to high accuracy. 
This is also confirmed in the calculations of ${A ^{2}\Pi_{1/2}} \leftrightarrow X^2\Sigma^{+}$ transition dipole moments; as shown in Table \ref{f13tms},
the computed ${A ^{2}\Pi_{1/2}} \leftrightarrow X^2\Sigma^{+}$ transition dipole moment takes significantly smaller values at 2.04 {\AA} and 2.06 {\AA} than the neighbouring bond lengths due to mixing of the $4f^{-1}_{5/2,1/2}$ state.

An alternative excitation scheme to obtain the $X^2\Sigma^{+}$ and $A ^{2}\Pi_{1/2}$ states is 
to use the closed-shell cationic state that corresponds to Yb atomic configuration $4f^{14}6s^06p^0$ as the reference state and then attach one electron to the $6s$ or $6p_{1/2}$ orbital. 
Since the cationic state is a well-defined single-reference state, this scheme 
can provide an accurate description for the $X^2\Sigma^{+}$ and $A ^{2}\Pi_{1/2}$ states. 
Our present results agree well with the EOM-CCSD results in Ref. \cite{Pototschnig21} obtained using this scheme. On the other hand, within this exciation scheme the $4f^{-1}_{J,|\Omega|}$ states have to be obtained using a ``double excitation" involving an electron attachment to one 6s orbital together with an excitation from a 4f orbital to the other 6s orbital. Since in general EOM-CCSD cannot provide accurate treatments for doubly excited states, this scheme in combination with EOM-CCSD cannot provide accurate treatments for the $4f^{-1}_{J,|\Omega|}$ states. 
Another possible excitation scheme is to use the anionic state 
that corresponds to Yb atomic configuration $4f^{14}6s^{2}$ as the reference state and 
to obtain the ground state by removing one 6s electron and a $4f^{-1}_{J,|\Omega|}$ state by removing a 4f electron. However, the anionic state exhibits large double amplitudes in CCSD calculations; it is not a well-defined single reference state. EOM-CCSD calculations using the anionic state as the reference thus exhibit relatively large errors, e.g., we have found that EOM-CCSD and EOM-CCSDT energies for the $4f^{-1}_{J,|\Omega|}$ states obtained using this scheme differ by more than 0.5 eV. While the EOM-CCSD results in Ref. \cite{Pototschnig21} obtained using this scheme
show significant overestimation of the energies for the $4f^{-1}_{J,|\Omega|}$ states, 
the relative energies within the $4f^{-1}_{J,|\Omega|}$ manifold agree well with our present calculations. 
Nevertheless, 
the present excitation scheme as summarized in Table \ref{excite} treats $4f^{-1}_{5/2,1/2}$ and $A ^{2}\Pi_{1/2}$ states using single excitations on the same reference state and thus enables accurate EOM-CCSD calculations of transition moments between these states. This is an ability that the other excitation schemes cannot offer and renders the present scheme the method of choice.

The use of an open-shell reference function in the present EOM-CCSD calculations 
leads to time-reversal symmetry breaking. For example, the absolute electronic energies for the $4f^{-1}_{7/2,1/2}, \Omega=1/2$ state and the $4f^{-1}_{7/2,1/2}, \Omega=-1/2$ state obtained in our calculations differ by around 50 cm$^{-1}$. The computed energies for the $A ^{2}\Pi_{1/2}, \Omega=-1/2$ states 
are about 600 cm$^{-1}$ lower than those of the $A ^{2}\Pi_{1/2}, \Omega=1/2$ states. 
Interestingly, since the computed term energy of the $A^2\Pi_{1/2}, \Omega=1/2$ state as shown in Table \ref{f13ene} is 
around 200 cm$^{-1}$ higher than the experimental value,
the average of the computed term energies of the $A^2\Pi_{1/2}, \Omega=\pm1/2$ states 
are close to the experimental value.
The time-reversal symmetry breaking is a non-trivial issue for EOM calculations using
open-shell reference functions, \change{although such calculations
have been shown to be accurate as long as the reference state is
single reference \cite{Comeau93,Levchenko03,Pathak16}}.
%This requires further investigation, 
In the present study, this issue does not affect the main conclusion but merits further investigation.

\subsection{Branching ratios for the $A ^{2}\Pi_{1/2}\rightarrow 4f^{-1}_{7/2,1/2}$ transitions
and lifetime of the $4f^{-1}_{7/2,1/2}$ state}

\begin {table*}
\caption {The square of transition dipole moments (in units of $(e a_0)^2$) obtained
from X2CAMF-CCSD/QZ calculations. }
\begin{center}
\begin{tabular}{c|ccc} 
\hline
 ~ $R_e$ ~ & ~ $A ^{2}\Pi_{1/2}\leftrightarrow 4f^{-1}_{7/2,1/2}$ ~ &
 ~ ${A ^{2}\Pi_{1/2}} \leftrightarrow X^2\Sigma^{+}$ ~
 & ~ $4f^{-1}_{7/2,1/2} \leftrightarrow X^2\Sigma^{+}$ ~ \\
 %& ~ $X^2\Sigma_{1/2}\leftrightarrow 4f^{-1}_{7/2,3/2}$ ~ \\
\hline
1.88 & 0.0418 & 3.48       &1.00E-3 \\ % & 9.51E-4 \\ %& 4.71E-4 \\
1.90 & 0.0359 & 3.60       &6.90E-4 \\ % & 6.54E-4 \\ %& 4.29E-4 \\
1.92 & 0.0296 & 3.68       &4.66E-4 \\ % & 4.40E-4 \\ %& 3.89E-4 \\
1.94 & 0.0254 & 3.74       &3.05E-4 \\ % & 2.87E-4 \\ %& 3.52E-4 \\
1.96 & 0.0212 & 3.77       &1.92E-4 \\ % & 1.78E-4 \\ %& 3.17E-4 \\
1.98 & 0.0170 & 3.79       &1.13E-4 \\ %    & 1.02E-4 \\ %& 2.85E-4 \\
2.00 & 0.0149 & 3.80       &0.58E-4 \\ % & 0.49E-4 \\ %& 2.55E-4 \\
2.02 & 0.0118 & 3.77       &0.21E-4 \\ % & 0.14E-4 \\ %& 2.28E-4 \\
2.04 & 0.0100 & 3.56       &0.02E-4 \\ % & 0.07E-4 \\ %& 2.03E-4 \\
2.06 & 0.0082 & 3.49       &0.16E-4 \\ % & 0.20E-4 \\ %& 1.80E-4 \\
2.08 & 0.0072 & 3.80       &0.24E-4 \\ % & 0.27E-4 \\ %& 1.59E-4 \\
%2.04 & 0.0100 & 2.03 (3.83) & 0.07E-4 \\ %& 2.03E-4 \\
%2.06 & 0.0082 & 3.79 (3.85) & 0.20E-4 \\ %& 1.80E-4 \\
%2.08 & 0.0072 & 3.85 (3.85) & 0.27E-4 \\ %& 1.59E-4 \\
%1.88 & 0.0410 & 3.54       &1.00E-3 \\ % & 9.51E-4 \\ %& 4.71E-4 \\
%1.90 & 0.0354 & 3.65       &6.90E-4 \\ % & 6.54E-4 \\ %& 4.29E-4 \\
%1.92 & 0.0296 & 3.73       &4.66E-4 \\ % & 4.40E-4 \\ %& 3.89E-4 \\
%1.94 & 0.0247 & 3.79       &3.05E-4 \\ % & 2.87E-4 \\ %& 3.52E-4 \\
%1.96 & 0.0203 & 3.83       &1.92E-4 \\ % & 1.78E-4 \\ %& 3.17E-4 \\
%1.98 & 0.0170 & 3.84       &1.13E-4 \\ %    & 1.02E-4 \\ %& 2.85E-4 \\
%2.00 & 0.0144 & 3.83       &0.58E-4 \\ % & 0.49E-4 \\ %& 2.55E-4 \\
%2.02 & 0.0118 & 3.73       &0.21E-4 \\ % & 0.14E-4 \\ %& 2.28E-4 \\
%2.04 & 0.0100 & 3.83       &0.02E-4 \\ % & 0.07E-4 \\ %& 2.03E-4 \\
%2.06 & 0.0082 & 3.79       &0.16E-4 \\ % & 0.20E-4 \\ %& 1.80E-4 \\
%2.08 & 0.0072 & 3.85       &0.24E-4 \\ % & 0.27E-4 \\ %& 1.59E-4 \\
\hline
\end{tabular}
\label{f13tms}
\end{center}
%a. fixed to the value derived in the previous subsection. \\
%b. small discontinuity in the surface because of perturbation by another state. 
\end {table*}

The spontaneous decay rates for the $A ^{2}\Pi_{1/2}\rightarrow 4f^{-1}_{7/2,1/2}$ transitions relative to
those of the $A ^{2}\Pi_{1/2}\rightarrow X^2\Sigma^+$ transitions determine the loss of population to the $4f^{-1}_{7/2,1/2}$ states during laser cooling on the $A ^{2}\Pi_{1/2}\leftrightarrow X^2\Sigma^{+}$ optical cycling transition. We have calculated the transition dipole moments 
for these transitions using the X2CAMF-EOM-CCSD wave functions of 
the $X^2\Sigma^+$, $A ^{2}\Pi_{1/2}$, and $4f^{-1}_{7/2,1/2}$ states. 
The values for the square of transition dipole moments in the range of the Yb-F bond length of 1.88-2.08 {\AA} 
are summarized in Table \ref{f13tms}. 
The $A ^{2}\Pi_{1/2}\leftrightarrow 4f^{-1}_{7/2,1/2}$ transition dipole moments 
amount to a few to ten percent those of the $A ^{2}\Pi_{1/2}\leftrightarrow X^2\Sigma^{+}$ transitions. 
For example, for a Yb-F bond length of 2.00 {\AA},
the $A ^{2}\Pi_{1/2}\leftrightarrow 4f^{-1}_{7/2,1/2}$ transition dipole moment is around 6\% of the $A ^{2}\Pi_{1/2}\leftrightarrow X^2\Sigma^{+}$ transition dipole moment. We note that the computed value of $1.96 e a_0$ for the $ A ^{2}\Pi_{1/2} \leftrightarrow X^{2}\Sigma^{+}$ transition dipole moment at the equilibrium $R_e$ is within 13\% of the measured value of $(1.73\pm 0.06)e a_0$~\cite{Zhuang2011}.

The $A ^{2}\Pi_{1/2}\leftrightarrow 4f^{-1}_{7/2,1/2}$ transitions are nominally
two-electron processes. A close inspection of the EOM-CCSD wave functions reveals that
the dominating mechanism for borrowing transition strength for these nominally forbidden transitions
is the mixing of the $X^2\Sigma^{+}$ configuration into the wave function of the $4f^{-1}_{7/2,1/2}$ states. For example, at a Yb-F bond length of 2.00 {\AA}, the coefficient of the ground-state determinant in the X2CAMF-EOM-CCSD/QZ wave function for the $4f^{-1}_{7/2,1/2}$ states is around 0.054.
Based on simple perturbation theory, the mixing coefficient should be roughly inverse proportional to the energy difference between the $4f^{-1}_{7/2,1/2}$ and $X^2\Sigma^{+}$ state. The X2CAMF-EOM-CCSD/TZ value of around 5500 cm$^{-1}$ for the energy difference and 0.062 for the mixing coefficient are consistent with the QZ value of around 6200 cm$^{-1}$ for the energy difference and 0.054 for the coefficient. As shown in Table \ref{f13ene}, the estimated energy difference is around 1800 cm$^{-1}$ higher than the QZ value. If we use a value of around 8000 cm$^{-1}$ for this energy difference at the Yb-F bond length of 2.00 {\AA}, the mixing coefficient is estimated to be around 0.042. The transition intensities obtained using X2CAMF-EOM-CCSD/QZ transition dipole moments thus are qualitatively correct with perhaps an overestimation of around 40\%. Note that shorter bond lengths exhibit smaller gaps between $4f^{-1}_{7/2,1/2}$ and $X^2\Sigma^{+}$ states and hence larger $A ^{2}\Pi_{1/2}\leftrightarrow 4f^{-1}_{7/2,1/2}$ transition dipole moments. 
We also mention that the $A ^{2}\Pi_{1/2}\leftrightarrow 4f^{-1}_{7/2,3/2}$ transition dipole moments are one order of magnitude smaller than those of $A ^{2}\Pi_{1/2}\leftrightarrow 4f^{-1}_{7/2,1/2}$ transitions, since the $4f^{-1}_{7/2,3/2}$ states do not have contributions from the ground-state configuration. The transition intensities of the $A ^{2}\Pi_{1/2}\leftrightarrow 4f^{-1}_{7/2,3/2}$ 
transitions are thus two orders of magnitude lower and do not play a significant role in the present discussions. \change{We also find that the transition dipole moments for the $4f^{-1}_{5/2,1/2} \rightarrow 4f^{-1}_{7/2,1/2}$ transitions are two orders of magnitude smaller than those of the $A ^{2}\Pi_{1/2}\rightarrow X^2\Sigma^+$ transitions. Consequently, mixing of the $A ^{2}\Pi_{1/2}$ and $4f^{-1}_{5/2,1/2}$ states, together with $4f^{-1}_{5/2,1/2} \rightarrow 4f^{-1}_{7/2,1/2}$ transitions, do not play an important role. }

Based on the computed transition dipole moments and the energy level positions derived in Table \ref{f13ene}, 
the $A ^{2}\Pi_{1/2}$ to $4f^{-1}_{7/2,1/2}$ spontaneous decay pathways
have branching ratios of around 0.05\% compared with the $A ^{2}\Pi_{1/2}\rightarrow X^2\Sigma^{+}$ decay pathways
and thus become relevant when scattering around 1,000-2,000 photons. 
We have also performed DVR calculations for vibronic branching ratios and summarized the results in Table \ref{theory_branching}. 
The branching ratios for the $A ^{2}\Pi_{1/2}(v=0) \rightarrow 4f^{-1}_{7/2,1/2} (v=0)$ and $A ^{2}\Pi_{1/2}(v=0) \rightarrow 4f^{-1}_{7/2,1/2} (v=1)$ transitions amount to 0.042\% and 0.010\%, respectively. They are comparable to that for 
the $A ^{2}\Pi_{1/2}(v=0) \rightarrow X^2\Sigma^{+} (v=3)$ transition.

Using the transition dipole moment near the equilibrium bond length for the $4f^{-1}_{7/2,1/2} \leftrightarrow X^2\Sigma^{+}$ transition presented in Table~\ref{f13tms}, and the corresponding transition frequency given in Table~\ref{f13ene}, we estimate the lifetime of the $4f^{-1}_{7/2,1/2}$ state to be approximately 8~ms.

\begin {table*}
\caption {
The $A ^{2}\Pi_{1/2}\rightarrow 4f^{-1}_{7/2,1/2}$ and $A ^{2}\Pi_{1/2}\rightarrow X^2\Sigma_{1/2}$ vibrational transitions with computed branching ratios greater than
0.001\%. Morse potentials defined using TZ spectroscopic constants in Table V and transition dipole strengths in Table VI were used in the DVR calculations. }
\label{theory_branching} 
\begin{center}
\begin{tabular}{c|cc} 
\hline
 ~ Transitions ~ & ~ Branching ratios ~ & Closure\\
\hline
 ~ $A ^{2}\Pi_{1/2}(v=0) \rightarrow X^2\Sigma^+ (v=0)$ ~  & 93.178\%  & 93.178\% \\
  $A ^{2}\Pi_{1/2}(v=0) \rightarrow X^2\Sigma^+ (v=1)$  & 6.474\% & 99.652\% \\
$A ^{2}\Pi_{1/2}(v=0) \rightarrow X^2\Sigma^+ (v=2)$  & 0.284\% & 99.937\% \\
$A ^{2}\Pi_{1/2}(v=0) \rightarrow 4f^{-1}_{7/2,1/2} (v=0)$  & 0.042\% & 99.979\%\\
$A ^{2}\Pi_{1/2}(v=0) \rightarrow X^2\Sigma^+ (v=3)$  & 0.010\% & 99.989\%\\
$A ^{2}\Pi_{1/2}(v=0) \rightarrow 4f^{-1}_{7/2,1/2} (v=1)$  & 0.010\% & 99.999\%\\
%$A ^{2}\Pi_{1/2}(v=0) \rightarrow X^2\Sigma^+ (v=4)$  & 0.0003\% & 99.997\%\\
$A ^{2}\Pi_{1/2}(v=0) \rightarrow 4f^{-1}_{7/2,1/2} (v=2)$  & 0.001\% & $>$ 99.999\%\\
% ~ $A ^{2}\Pi_{1/2}(v=0) \rightarrow X^2\Sigma^+ (v=0)$ ~  & 91.702\%  & 91.702\% \\
%  $A ^{2}\Pi_{1/2}(v=0) \rightarrow X^2\Sigma^+ (v=1)$  & 7.725\% & 99.428\% \\
%$A ^{2}\Pi_{1/2}(v=0) \rightarrow X^2\Sigma^+ (v=2)$  & 0.475\% & 99.903\% \\
%$A ^{2}\Pi_{1/2}(v=0) \rightarrow 4f^{-1}_{7/2,1/2} (v=0)$  & 0.045\% & 99.948\%\\
%$A ^{2}\Pi_{1/2}(v=0) \rightarrow X^2\Sigma^+ (v=3)$  & 0.034\% & 99.981\%\\
%$A ^{2}\Pi_{1/2}(v=0) \rightarrow 4f^{-1}_{7/2,1/2} (v=1)$  & 0.013\% & 99.994\%\\
%$A ^{2}\Pi_{1/2}(v=0) \rightarrow X^2\Sigma^+ (v=4)$  & 0.003\% & 99.997\%\\
%$A ^{2}\Pi_{1/2}(v=0) \rightarrow 4f^{-1}_{7/2,1/2} (v=2)$  & 0.002\% & 99.999\%\\
\hline
\end{tabular}
\end{center}
%a. fixed to the value derived in the previous subsection. \\
%b. small discontinuity in the surface because of perturbation by another state. 
\end {table*}

\section{Summary and discussion}
\label{summary}

In this paper, we have carried out a de-perturbation analysis of spectroscopic data and {\it ab initio} electronic structure calculations to obtain useful information about low-lying electronic states of YbF arising from excitation of an inner-shell, 4f electron. The de-perturbation analysis determines the energy, bond length and vibrational constants ($\omega_e$ and $\omega_e x_e$) of the  $4f^{-1}_{5/2,1/2}$ state to high precision. \change{The bond lengths and vibrational constants are in very good agreement with those obtained from our electronic structure calculations}, validating the accuracy of those calculations.  A coupling of 108.3~cm$^{-1}$ between the A$^{2}\Pi_{1/2}$ and $4f^{-1}_{5/2,1/2}$ states results in a set of vibrational eigenstates of mixed character. Even the lowest vibrational level of the set, previously considered to be a pure A$^{2}\Pi_{1/2} (v=0)$ state, contains a significant admixture of $4f^{-1}_{5/2,1/2}$. These mixed states are used in the scheme designed for laser cooling of YbF, so establishing their exact nature is important. Notably, our analysis establishes the radiative branching ratios from the mixed states to the vibrational levels of $X^{2}\Sigma^+$. These branching ratios agree well with measurements, and the analysis explains why they take the values they do, which was not previously understood. The electronic structure calculations determine the energies and spectroscopic constants of all 7 of the $4f^{-1}_{5/2}$ and $4f^{-1}_{7/2}$ states, the branching ratios for the $A ^{2}\Pi_{1/2} \rightarrow 4f^{-1}_{7/2,1/2}$ transitions and the lifetime of the $4f^{-1}_{7/2,1/2}$ state.

Decay from $A^{2}\Pi_{1/2}$ to $4f^{-1}_{7/2,1/2}$ is a concern for laser cooling, even at the $10^{-4}$ or $10^{-5}$ level, because it represents a leak out of the cooling cycle. We were initially concerned that such a leak would open up due to the very strong mixing of $A^{2}\Pi_{1/2} (v=1)$ with $4f^{-1}_{5/2,1/2}(v=0)$, together with a small, but not insignificant, strength of $4f^{-1}_{5/2,1/2} \rightarrow 4f^{-1}_{7/2,1/2}$ transitions. However, since f-f transitions are nominally dipole forbidden, 
our calculations find the electronic transition dipole moment of $4f^{-1}_{5/2,1/2} \rightarrow 4f^{-1}_{7/2,1/2}$ transitions to be two orders of magnitude smaller than those of $A ^{2}\Pi_{1/2}\rightarrow X^2\Sigma^+$ transitions, and thus too weak to be of concern. Instead, a leak is opened up by a non-negligible mixing of $4f^{-1}_{7/2,1/2}$ with the leading configuration of the ground $X ^{2}\Sigma^{+}$ state together with the very strong $A^{2}\Pi_{1/2} \leftrightarrow X ^{2}\Sigma^{+}$ transition dipole moment. In other words, dynamic correlation effects that introduce
the $4f^{14}6s^1$ configuration into the many-electron wave function of the $4f^{-1}_{7/2,1/2}$ state is responsible for this mechanism. The leak to the $4f^{-1}_{7/2,1/2}$ states has a branching ratio of about 0.05\%, which limits the number of photons that can be scattered to roughly 2000. This is insufficient for laser slowing and magneto-optical trapping. For example, 2000 photons scattered from counter-propagating laser light only reduces the velocity of a YbF molecule by 7.4~m/s. 

One option to address this problem is simply to wait for the $4f^{-1}_{7/2,1/2}$ states to spontaneously decay to the $X^{2}\Sigma^{+}$ states. This takes the molecules to rotational states of the wrong parity for laser cooling, but the population in these states can be returned to the cooling cycle by coupling the rotational states together using a microwave field. Such microwave re-mixing has been applied previously to YO molecules to eliminate a leak out of the cooling cycle arising from a weak decay to an intermediate $^{2}\Delta_{3/2}$ state, leading to successful laser slowing and trapping of these molecules~\cite{Yeo2015, Collopy2018}. In the case of YO, the leak is approximately the same size as found here for YbF. However, in YO the lifetime of the intermediate $^{2}\Delta_{3/2}$ state is $23\pm 2$~$\mu$s~\cite{Zhang(2)2020}, so population returns rapidly to the ground state. The much longer lifetime of about 8~ms that we find here for the $4f^{-1}_{7/2,1/2}$ state would limit the time-averaged scattering rate to about $2\times 10^{5}$ photons per second, which is an order of magnitude slower than could otherwise be achieved. For this reason, microwave re-mixing is not a good solution to the problem of the leak.

A second option is to pump the population that reaches the $4f^{-1}_{7/2,1/2}$ states back to $A^{2}\Pi_{1/2}$, or to a higher-lying state with a strong decay channel to $X^{2}\Sigma^{+}$. Fortunately, the leak only populates a few vibrational states. We see from Table~\ref{theory_branching} that by addressing just the two lowest vibrational states of $4f^{-1}_{7/2,1/2}$ the number of scattered photons can be increased above $10^{4}$. This does not add too much extra complexity in terms of the number of lasers needed. To avoid stimulating signifcant population into these states, the repump lasers should not connect to $A^{2}\Pi_{1/2}(v=0)$, but instead to one of the higher lying vibrational states such as the mixed states [557] or [561]. The wavelength is close to 1~$\mu$m, a very convenient wavelength for laser technology. At present, the main hindrance to developing a suitable repumping scheme is the large uncertainty (approx 200~cm$^{-1}$) in the energies of the $4f^{-1}_{7/2,1/2}$ states and the lack of any spectroscopic data on these levels. Thus, a spectroscopic characterization of the $4f^{-1}_{7/2,1/2}$ levels is urgently needed. A possible starting point is a spectral analysis of the fluorescence emitted on the $A ^{2}\Pi_{1/2} \rightarrow 4f^{-1}_{7/2,1/2}$ transition, but this is difficult because the transition has only a small branching ratio and the emission near 1~$\mu$m is much more difficult to detect than emission in the visible. An alternative might be a broadband excitation (e.g using a pulsed dye laser) from the $A^{2}\Pi_{1/2}$ state to higher lying states that have $4f^{13}$ character, such as states correlating to Yb$^+$ $4f^{13}6s6p$ followed by a spectral analysis of the resulting fluorescence. States of this kind are likely to have strong emission in the visible to $4f^{-1}_{7/2,1/2}$.

Finally, we note that there are deceleration techniques that avoid photon scattering altogether, or only require a small number of photons to be scattered. These include Stark deceleration which has recently been used to decelerate quite heavy molecules~\cite{Aggarwal2021}, Zeeman Sisyphus deceleration~\cite{Fitch2016} recently demonstrated for polyatomic CaOH~\cite{Augenbraun2021}, and centrifuge deceleration~\cite{Wu2017}. If it proves difficult to close the leak from the cooling cycle, these slowing methods could be employed to bring molecules to rest. At that point the molecules can be cooled very rapidly and efficiently to low temperature using an optical molasses, which requires only a few hundred photon scattering events to be effective and has already been demonstrated in one and two dimensions for YbF~\cite{Lim2018, Alauze2021}.

\begin{acknowledgments}

We are grateful to Jongseok Lim, Xavier Alauze, Anastsia Borschevsky, Pi Haase, Luk\'a\v{s} F\'elix Pa\v{s}teka and Johann Pototschnig for helpful discussions on the topic of this paper.

The work at Imperial College London has received support from the UK Science and Technology Facilities Council (ST/S000011/1), the European Commission (101018992), the John Templeton Foundation (grant 61104), the Sloan Foundation (G-2019-12505) and the Gordon and Betty Moore Foundation (grant 8864). The computational work at the Johns Hopkins University by C. Z. and L. C. was supported by the National Science Foundation, under Grant No. PHY-2011794. 
Computations at Johns Hopkins University were carried out
at Advanced Research Computing at Hopkins (ARCH) core facility (rockfish.jhu.edu), which
is supported by the National Science Foundation (NSF) under grant number OAC-1920103.

\end{acknowledgments}

\bibliography{references,soeomcc}

\end{document}